# A flexible GPU-accelerated radio-frequency readout for superconducting detectors

L. Minutolo, B. Steinbach, A. Wandui, and R. O'Brient

*Abstract*— We have developed a flexible radio-frequency readout system suitable for a variety of superconducting detectors commonly used in millimeter and submillimeter astrophysics, including Kinetic Inductance detectors (KIDs), Thermal KID bolometers (TKIDs), and Quantum Capacitance Detectors (QCDs). Our system avoids custom FPGA-based readouts and instead uses commercially available software radio hardware for ADC/DAC and a GPU to handle real time signal processing. Because this system is written in common C++/CUDA, the range of different algorithms that can be quickly implemented make it suitable for the readout of many others cryogenic detectors and for the testing of different and possibly more effective data acquisition schemes.

*Index Terms*—superconducting detector readout, Kinetic inductance detector, GPU computing, demodulation algorithm, polyphase Filter Bank.

## I. INTRODUCTION

SUB-MILLIMETER detectors such as KIDs, TKIDs, and QCDs use high-Q radio-frequency resonances to accomplish dense frequency domain multiplexing. Warm multiplexed readout electronics has proven to be a bottleneck in pushing these detectors from single-prototype to larger arrays. Traditional systems use FPGA firmware which requires expertise to modify and develop. As an example, ROACH boards make use of Xilinx FPGAs to real-time process the raw data acquired by ADC(s) and generate the buffers fed to the DAC(s) [1,2]. While FPGA centered readouts can operate at low powers and with margin on speed for many applications, the multi-month development time for their firmware is slow and requires the expense of employing a full-time expert.

An alternative is to off-load the computationally demanding tasks to a common desktop computer, accelerating sub-tasks in the computer's Graphical Processing Unit (GPU) as is appropriate. This choice allows us to use Nvidia's CUDA, a public library of C++ compatible functions widely used for developing GPU applications in graphics, AI, and big data processing. The development time for the required Python and C++ software is reduced to days, allowing quick adjustments of the readout system to meet the needs of different detector programs. Moreover, most scientists can write and edit C++ and Python software, obviating the expert needed to maintain an FPGA system.

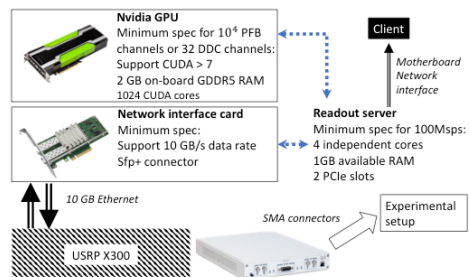

Fig. 1 Flow of the data flow through the hardware with minimum specs for each component. The Nvidia GPU is represented as a single block but can be replaced by as many cards as the system can support.

## II. SYSTEM OVERVIEW

As shown in summarized in Fig. 1, the readout system is composed exclusively of commercial off-the-shelf (COTS) components. This reduces the cost of reproducing the system and allows us to use open source libraries of functions to co-ordinate the hardware.

We use the following components:
- X300 Ettus research Software Defined Radio (SDR)
- Intel X520 10-Gbit Ethernet card and cables.
- Nvidia GTX-1080 Ti GPU (GDDR5X)

While all hardware is supported by Linux-compatible drivers, the drivers and libraries are cross-platform, which has let us run this system on Windows and macOS operating systems. The performance of the readout varies slightly between different platforms because of differing OS scheduler behavior. Our test platform consists of Ubuntu 18.04 using the g++ compiler 6.4, CUDA 9.1 [3], UHD 3.13 [4] and boost 1.64 [5].

### A. Ettus Universal Software Radio Peripheral (USRP) SRD

The X-300 is an SDR produced by Ettus Research (National Instruments). The X-300 motherboard coordinates two 14-bit ADCs and 16-bit DACs with a Xilinx Kintex-7 FPGA. This FPGA layer is computationally "thin," avoiding any high level processing beyond its basic task of wave-form generation and data digitization. The system streams data to/from an external computer via 10-GBit Ethernet at a rate of 200Ms/s, matching the ADC sample rate.

The motherboard interfaces with pairs of daughterboards that can mix from baseband into microwave tones as high as

6GHz, using an internally generated LO for mixing. Most of our team's work has used the:
- WBX-120: mixes to 20-2200MHz range.
- SBX-120: mixes to 400-4400MHz range.

Each card provides 120MHz of RF bandwidth, so 240MHz of theoretical bandwidth between the two slots.

Ettus maintains open-source python and C++ libraries, called UHD libraries [4], to utilize this hardware as parts of custom systems.

### B. Intel 10-Gbit Ethernet Card

The computer and USRP communicate via 10-GBit ethernet coordinated with an Intel X520 Converged Network Adapter, which delivers data to 8 lanes of the computer's PCI Express 2.0 bus. All Linux operating systems have drivers to work with this hardware. As shown in Fig 1, the incoming data transmits to the computer's RAM before being copied to the GPU's onboard RAM. Outgoing data follows the reverse path during tone generation. These steps currently limit the system total speed.

### C. Nvidia GPU

We can process data in the server's CPU for limited bandwidth applications (one or two detectors) or the GPU when working with larger sample counts. The CPU/GPU is responsible for channelizing or demodulating, filtering, and decimating. We use a GTX-1080 GPU, which contains 2560 cores capable of supporting the 2560 parallel threads.

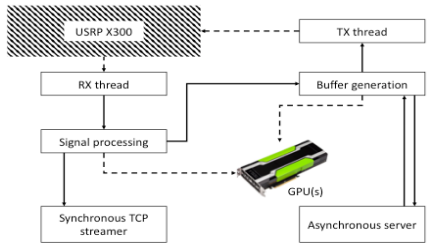

Fig. 2 Partitioning of tasks into threads and the common pattern used to exchange data and metadata. Each solid line represents a consumer/producer relation while the dashed lines represent a PCIe/Ethernet link.

### III. SOFTWARE

Our software must coordinate a diverse set of tasks, potentially spread across many computers with the requirement that it port to different operating systems. To handle these challenges, our readout system CPU spawns various threads (Fig. 2):

1. Transmission of buffers to USRP (Tx thread)
2. Reception of data from USRP (Rx thread)
3. Transmission buffer producer
4. Rx signal processing
5. Streaming of processed data via TCP protocol
6. Asynchronous server thread
7. Synchronous TCP streamer

All threads share buffers and metadata via lock-free queues of pointers to pre-allocated memory regions. The threads managing the queues and the asynchronous server operations are based on the Boost [5] collection of C++ functions.

Both the processing of the acquired signal and the generation of the transmission buffer occur in the GPU. The role of the handling thread is only to tune the performance of the GPU kernels to optimally upload/download data to/from the GPU and to handle buffer metadata. In applications where we use active feed-back, the buffer generator checks for eventual input from the signal processing thread and modifies the output buffer kernel's generator consequently. The data processed with the GPU are then sent to a client interface via TCP protocol.

CUDA can help parallelize computations in the GPU if they are expressed as matrix computations. The following subsections explain our development with this principle in mind.

### A. Polyphase Filter Bank

Like the FPGA based ROACH boards, the GPU system channelizes with a polyphase filter bank (PFB). The PFB controls the frequency response of the FFT bins by pre-filtering raw data in a finite impulse response (FIR) filter [6,7]. In our case, the FIR filter is a product of a sinc and a hamming function of length $N \cdot A_v$, where N is the decimation factor and $A_v$ is the filter taps per frequency channel. We construct the sinc function such that its spectral response has a cut-off frequency that controls cross-talk of adjacent FFT bins to be less than 6dB. Figure 3 shows the cross-talk levels in our implementation.

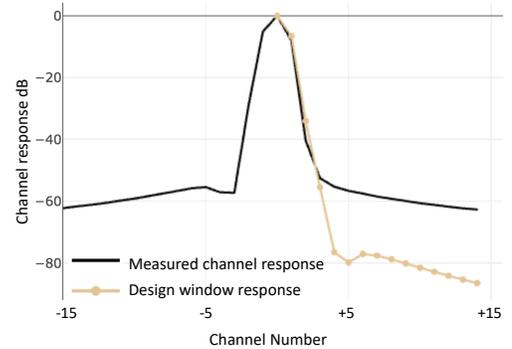

Fig 3. The spectral response of a single tone of 418MHz fed into the USRP RX port and channelized with decimation of $N=10^5$ and $A_v=4$. Cross-talk in adjacent channels is approximately 6dB, as per design.

The PFB is intrinsically parallel friendly because it relies on processing different chunks of raw data independently and then summing. In our case, we multiply $N \cdot A_v$ samples long sections of data against the FIR filter, but across $N$ GPU cores run in parallel. This effectively reformats the data as a $N \times A_v$ matrix (or what CUDA terms a "texture"). We transpose and multiply against a unit vector of magnitude $1/N$ and length $A_v$ to decimate. All of these steps are efficiently handled by functions in the cuBLAS library [8].

Once the buffer has been filtered and decimated, we batch call cuFFT [9] to fourier transform each section of data. We have used our system to channelize signals into as many as $10^7$ bins, and the system can return the full spectrum to the user or specific user defined channels.

### B. Direct Demodulation

As an alternative to channelization, we have used our system to directly demodulate the time-stream against sets of specific user defined tones. We multiply a sample of raw time-stream data against buffers of different frequency tones in parallel in different cores of the GPU. We control data volume with decimation and anti-aliasing filters (moving average of a square window). These filtering and decimation steps map to the same cuBLAS matrix multiplication as for the PFB [7].

We have used our system to demodulate ~ 200 channels with a decimation factor of 500, sampled at 100Msps.

### C. Chirped readout

VNA scans such as in Figure 5 require the frequency to be swept over time with a quadratic change in phase:

$$F(t) = A \cdot \sin(\phi_t)$$
$$\phi_t = 2\pi ( f_0 t + k t^2 ) \quad \text{Eq. 1}$$

where $f_o$ is the initial frequency and $k$ the constant ramp rate (MHz/s).

Generating chirped waveforms is a memory intensive process- a 20s duration chirp sampled at 100MHz described with 32 bit precision would require 16GB of free memory. Instead, we parallelize this process in the GPU to generate excitations real time and therefore only have to allocate a single packet's memory in RAM. We can directly demodulate the Rx signal against this waveform using the algorithm outlined in section IIIB.

A potentially more powerful algorithm is a "fast chirp" which deposits energy in the resonators array by swiping over the entire bandwidth over the course of a few micro-seconds, far faster than the ring-down time of the resonators. After this short transmission phase, we turn off the excitation and record the Rx time-stream as the resonators ring down. We perform the FFT real time using the PFB algorithm described in section IIIA.

### D. Pulsed readout

We have performed experiments with our TKIDs using an excitation of a square pulse followed by a listen phase, similar to the fast chirp described in IIIC. Our goal was to look for ring-down effects that might be associated with two level systems (TLS). Had we seen a ring down, we could have subjected the fermionic TLS states to spin-echo spectroscopy where we would have followed the initial pulse with an inversion pulse a time $t$ later, and then searched for an echo an addition delay $t$ yet later. We did not see any ring-down from the pulse, likely because TLS lifetimes are ~1ns, far outside our system's bandwidth. But the fact that we could casually attempt this illustrates our readout's flexibility.

### IV. NOISE PERFORMANCE

We characterized the signal to noise ratio in our readout system with "loop-back" tests that replace the cryostat with a short coax cable. Fig. 4 shows such the results of such measurements, displaying spectra of the Signal to Noise ratio (S/N). We have removed a common mode component that produces high low frequency noise.

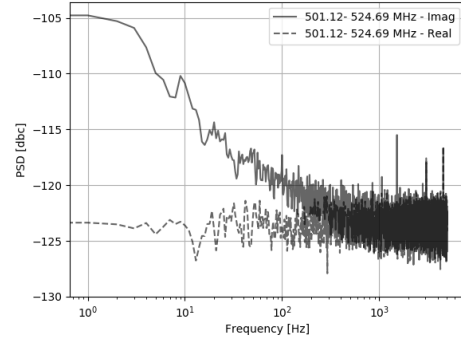

Fig. 4 Loopback Noise Measurements: Noise after removing a common mode component. We have also corrected by 14dB for shared power between channels.

We have performed these tests for single tones and combs of 10 tones and found that the S/N degrades by ~10dB as a result of nonlinear mixing of the different tones, likely in the ADC chips. We note that a chirped readout would be naturally immune to these effects.

Our TKIDs and KIDs optimized for atmospheric loading typically operate with ~80dBc signal to noise and the Low Noise amplifiers provide -97dBc, so this readout system provides ample sensitivity for characterizing them. The most demanding application we might use this system for is characterizing submillimeter spectrometers for flight, which need a S/N of 100-110dBc, still partially serviceable with this system.

### V. USE CASES

We illustrate the readout system's flexibility by showing two very different applications: beam mapping horn-coupled KIDs and photon statistics of QCDs.

#### A. Beam maps of horn-coupled KID array

JPL is developing a KID [10] aerial reconnaissance camera under contract from the US Navy, and we have tested our readout system with their cryostat. This cryostat has provided our team an opportunity to verify that our system can operate as a replacement to a conventional ROACH-2 system.

The focal plane is a monolithic horn coupled array of 128 lumped element KIDs, sensitive to a single polarization of 130-170GHz radiation. The cryostat has an HDPE window and a stack of low pass filters to limit loading on the focal plane.

The ROACH and GPU systems both detect 116 channels in the 100-200MHz spectral range on the readout line, as seen in VNA style measurements that slowly sweep the driving tone frequency to measure $S_{21}$. Each high-Q dip in Fig. 5a corresponds to a unique detector. We mapped the beams formed by the horns of all 116 detectors by using a pair of stepper motors to move a thermal source in front of the camera. Figure 5b shows one such beam. We note that the non

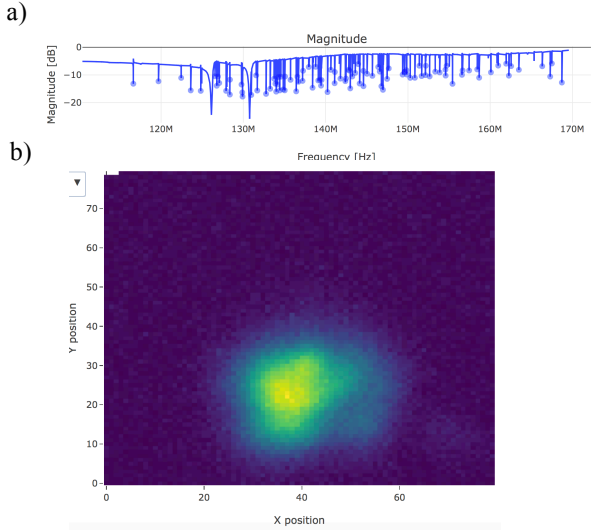

Fig. 5 (a) VNA scan of all KIDs in the Navy Cryostat. (b) Beam map of one detector out of the 88 maps collected. The visible Non-gaussian Structure was intrinsic to the camera.

-gaussian structure is present in both the GPU and ROACH read maps. It appears to be a consequence of the camera optics that is still in progress. This measurement demonstrated the flexibility of this system used for more complex measurements than VNA or noise studies and proves that it is stable over the course of the 1-2 hour duration of the map.

### B. Photon statistics in QCDs

JPL is developing QCDs for ultra-low background far-IR photon counting [11]. In the QCD, radiation couples to a superconducting absorber, breaking Cooper pairs in the absorber. This generates non-equilibrium quasiparticles, which diffuse to the junctions of a Single Cooper-pair Box (SCB), a device that features a bias dependent capacitance (the Quantum Capacitance). When quasiparticles tunnel across the SCB junctions, they change its average quantum capacitance, resulting in a shift in the center frequency of the resonator that we measure in the multiplexed readout. A photon absorption event generates 20 quasiparticles, triggering a burst of tunneling into the island that destroys a quantum capacitance peak. Missing peaks mark a single photon absorption event. Fig. 6a shows a time stream acquired with the GPU based readout. The data is processed by taking the standard deviation of chunks of the time stream spanning two quantum capacitance peaks and subtracting the results from its maximum. Figure 6b shows a histogram of the processed data for two illumination levels: $10^{-21}$W and $10^{-17}$W. The tall peaks at right are associated with the tunneling oscillations, whereas the small peaks at left are associated with disruptions generated by photon absorption.

### VI. FUTURE WORK

As mentioned above, the speed of our system is limited to the 2ms latency of writing data to and from the motherboard's RAM. SQUID readout of TESes cannot currently be serviced by our system because the electronics need to feedback at speeds fast compared to the electrothermally accelerated bolometer thermal time constant, typically 0.1-1ms. We are exploring how to use DMA compilers to write data directly from the Ethernet card to GPU without accessing the motherboard RAM, which may reduce the latency for this application.

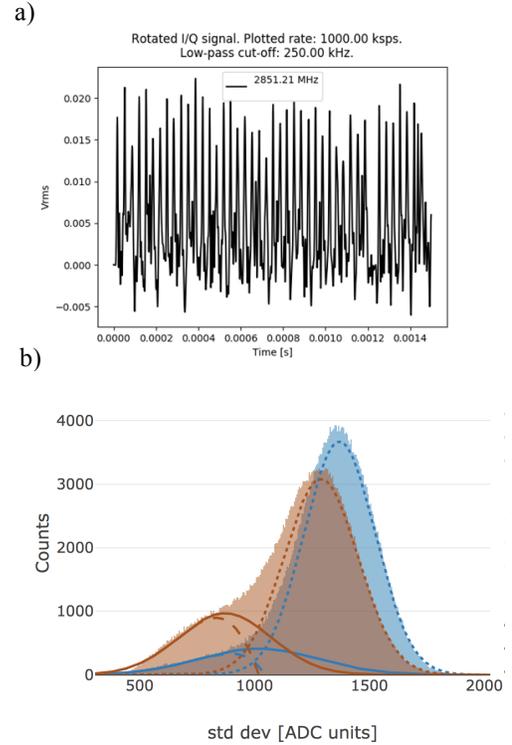

Fig 6. (a) QCD time-stream data taken with the GPU/USRP readout. There is a clear photon absorption at 0.0012s. (b) Histograms of deviations associated with oscillations seen in (a) The left peaks results from photon absorption, and the right from undisturbed quantum oscillations. The different curves correspond to different blackbody load temperatures- blue is $10^{-21}$W and orange $10^{-17}$W. The curves are fits to different components in the distributions.

### ACKNOWLEDGMENT

The research was carried out at the Jet Propulsion Laboratory, California Institute of Technology, under contract with the National Aeronautics and Space Administration. We thank Jonas Zmuidzinas, David Hawkins, and Ryan Monroe for useful discussions. We thank Peter Day, Daniel Cunnane and Jack Sayers for use of the JPL/Navy camera for testing as well as Pierre Echternach for use of his QCD testbed and prototypes. We acknowledge Atilla Kovacs' pioneering efforts in this GPU-centric approach.